\documentclass{aa} 
\usepackage{graphicx}
\usepackage{lscape}
\usepackage[varg]{txfonts}
\usepackage{amssymb}
\usepackage{stmaryrd}
\usepackage{url}
\usepackage{float} 
\usepackage{xspace}
\usepackage{siunitx}
\usepackage{textcomp} 
\usepackage{multirow}
\usepackage[usenames,dvipsnames]{xcolor}
\usepackage{longtable}
\usepackage{multicol}
\usepackage{tablefootnote}
\usepackage{footnote}
\makesavenoteenv{tabular}
\makesavenoteenv{table}
\usepackage{float}
\usepackage{rotating}
\usepackage{graphicx}
\usepackage{booktabs}
\usepackage{adjustbox}
\usepackage{natbib} 
\bibpunct{(}{)}{;}{a}{}{,}
\usepackage[colorlinks=true, linkcolor=blue, citecolor=blue, urlcolor=blue]{hyperref}
\newcommand{\citeyeartext}[1]{\citeauthor{#1} \citeyear{#1}}

\usepackage{amsmath}
\usepackage{tabularx}
\usepackage{colortbl}
\usepackage{stfloats}
\definecolor{lightgray}{gray}{0.9}

\newcommand{\rx}{RX\,J0122.9$-$7521\xspace}
\newcommand{\rxo}{\object{RX\,J0122.9$-$7521}\xspace}
\newcounter{Rco}
\newcommand{\Ionst}[1]{\setcounter{Rco}{#1}\Roman{Rco}}
\newcommand{\IonT}[2]{\mbox{#1\,{\scriptsize\MakeUppercase{\romannumeral #2}}}}
\newcommand{\Ion}[2]{\mbox{#1\,{\scriptsize\Ionst{#2}}}}
\newcommand{\Ionw}[3]{\mbox{#1\,{\scriptsize\Ionst{#2}}~$\lambda\,#3$\,\AA}}

\newcommand{\Ionww}[3]{\mbox{#1\,{\scriptsize\Ionst{#2}}~$\lambda\lambda\,#3$\,\AA}}

\newcommand{\se}[1]{\mbox{Sect.\,\ref{#1}}}
\newcommand{\logg}{\mbox{$\log g$}\xspace}
\newcommand{\loggw}[1]{\mbox{$\log g\hspace{-0.5mm} =\hspace{-0.5mm}  #1$}}

\newcommand{\Teff}{\mbox{$T_\mathrm{eff}$}\xspace}
\newcommand{\Teffw}[1]{\mbox{$\Teff\hspace{-0.5mm} =\hspace{-0.5mm} #1 \,\mathrm{kK}$}}
\newcommand{\Teffwo}[1]{\mbox{$\Teff\hspace{-0.5mm} =\hspace{-0.5mm} #1 \,\mathrm{K}$}}

\newcommand{\vrad}{$v_\mathrm{rad}$\xspace}

\newcommand{\Lsol}{$L_\odot$}
\newcommand{\Msol}{$M_\odot$}
\newcommand{\Rsol}{$R_\odot$}

\definecolor{greenlight}{rgb}{0.4, 0.8, 0.4}

\begin{document} 

\title{How an overweight and rapidly rotating PG\,1159 star in the Galactic halo challenges evolutionary models}
\titlerunning{Analysis of \rx}

\author{Nina\,Mackensen \inst{1}
    \and Nicole Reindl \inst{1}
    \and Klaus Werner \inst{2}
    \and Matti Dorsch \inst{3}
    \and Shuyu Tan \inst{4}
    }
    
\offprints{Nina\,Mackensen\\ \email{nmackensen@lsw.uni-heidelberg.de}}

\institute{Landessternwarte Heidelberg, Zentrum f\"ur Astronomie, Ruprecht-Karls-Universit\"at, Königstuhl~12, 69117 Heidelberg, Germany
\and Institut f\"ur Astronomie und Astrophysik, Kepler Center for Astro and Particle Physics, Eberhard Karls Universit\"at, Sand 1, 72076 T\"ubingen, Germany
\and Institut für Physik und Astronomie, Universität Potsdam, Haus 28, Karl-Liebknecht-Str. 24/25, 14476, Potsdam-Golm, Germany
\and Universität Heidelberg, Zentrum für Astronomie, Institut für Theoretische Astrophysik, Albert-Ueberle-Str. 2, 69120 Heidelberg, Germany}

\date{Received 19 March 2025 / Accepted 31 May 2025}

\abstract
{PG\,1159 stars are thought to be progenitors of the majority of H-deficient white dwarfs. Their unusual He-, C-, and O-dominated surface composition is typically believed to result from a late thermal pulse experienced by a single (pre-)white dwarf. Yet other formation channels -- involving close binary evolution -- have recently been proposed and could lead to similar surface compositions.\\
Here we present a non-local thermodynamic equilibrium spectral analysis based on new UV and archival optical spectra of one of the hottest PG\,1159 stars, \rxo. We find \Teffw{175} and \loggw{7.7}, and an astonishingly low O/C ratio of $7.3\times10^{-3}$ (by mass).
By combining the spectroscopic surface gravity and Gaia parallax with a 
spectral energy distribution fit, we derive a mass of $M_\mathrm{spec} = 1.8^{+1.1}_{-0.7} M_\odot$. Although this spectroscopic mass is higher than predicted by evolutionary models, it is subject to substantial 
uncertainty.
Furthermore, we find that \rx shows strongly rotationally broadened lines, suggesting that the previously reported photometric period of 41\,min indeed corresponds to the rotational period of this star. Our kinematic analysis shows that \rx belongs to the Galactic halo, which -- assuming single-star evolution -- is in stark contrast to its relatively high mass. The rapid rotation, high mass, and halo kinematics, as well as the lack of evidence of a close companion, led us to the belief that \rx formed through the merger of two white dwarfs. Yet, none of the current models can explain the surface abundances of \rx.
}


\maketitle
\newcommand{\uproman}[1]{\uppercase\expandafter{\romannumeral#1}}
\section{Introduction}
\label{sec:intro}
It is assumed that a star approaching the end of its life on the asymptotic giant branch (AGB) can undergo various evolutionary paths, not all of which lead to a transition to a hydrogen-rich white dwarf (DA). 
Thermal pulses (TPs) can occur not only during the AGB phase of a star's life but also during the post-AGB or white dwarf cooling phases. 
In the latter cases, these pulses and the associated dredge-up of helium-rich material from the inter-shell regions to the outer layers can cause the star to be temporarily thrown back onto the AGB and go through parts of it again \citep{Althaus_2009}. Depending on the timing of a TP, this process can result in a hydrogen-deficient white dwarf (DO), which are less common (20\%) than their hydrogen-rich counterparts \citep{DAundDO}.
Thus, it is assumed that a TP occurring during the departure from the AGB, but still during nuclear shell burning, i.e. a late thermal pulse (LTP), only results in a completely hydrogen-free atmosphere, making hydrogen temporarily undetectable but still leading to the formation of a DA \citep{Althaus_2005}. In contrast, a very late thermal pulse (VLTP) during the cooling phase of a white dwarf leads to an almost completely hydrogen-free atmosphere \citep{LTP1, schonberner1979asymptotic,LTB2}. Between the occurrence of TPs and the evolution into a DA or DO, stars likely pass through the extremely hot PG 1159 phase (\Teff = 60\,000 K - 250\,000 K). These stars are hydrogen-poor, due to either the dilution or the complete burning of atmospheric hydrogen, and their atmospheres are notably enriched in carbon and oxygen.
Over the last few decades, quantitative spectral analyses have revealed a considerable diversity in their elemental abundances (He = 0.30 - 0.92, C = 0.08 - 0.60, O = 0.02 - 0.20 mass fractions) and a large range of gravities (\(\log g\) = 5.3 - 8.3; see \citealt{Werner_2016} and, for the latest discoveries, \citealt{Werner_2024b}).
In addition, traces of hydrogen are occasionally detected. In such instances, they are classified as hybrid PG\,1159 stars. The existence of such a hybrid is currently being explained using the AGB final thermal pulse (AFTP) scenario \citep{Herwig2001}. In this scenario, a helium shell flash occurs precisely when the star is in the final stage of its AGB phase. This event influences the elemental abundance through simultaneous mixing and burning.
\\
However, single-star evolution through late He-shell flashes is not the only possible explanation of the observed surface abundances of PG\,1159 stars. 
A recently discussed possibility is a merger scenario in which PG\,1159 stars also originate from objects of the newly introduced spectral type CO-sdO. These CO-sdO stars are H-deficient, He-core-burning hot subdwarfs that exhibit unusually high abundances of C and O and are thought to result from the merger of a He-core white dwarf and a low-mass CO-core white dwarf (\citeyeartext{Werner+2022b}).
\begin{table*}[ht]
\centering
\caption{Values for the available spectra of \rx.}
\label{tab:obs}
\renewcommand{\arraystretch}{1.5}
\begin{tabular}{l l l l l l l} 
\hline
\hline
Telescope & Instrument & Grating & Resolving power& Coverage (\r{A}) & Exposure Time (s) & ProgID\\
\hline
HST & COS & G130M & 15\,000 - 19\,000 & 1133 - 1430 & 1967 & 17112 \\
\noalign{\smallskip}
VLT & UVES & CD2 | CD3 & 19\,540 | 18\,770 & {\fontsize{8.5}{10}\selectfont \renewcommand{\arraystretch}{0.8} \begin{tabular}[c]{@{}l@{}} 3280 - 4562 \\ 4582 - 6687 \end{tabular}} & 2 $\times$ 600 & 167.D-0407 \\
\noalign{\smallskip}
HST & STIS & G230LB | G430L | G750L & 1000 & 2000 - 10\,000 & 1400 & 14141 \\
\hline
\end{tabular}
\end{table*}
To date, approximately 70 PG\,1159 stars have been identified, 5 of which have been classified as hybrids. Recently, 29 of them, including \rx, were examined for pulsations by \cite{Sowicka2023}, as most are located within the GW Virginis instability strip.
A period of 41~min was found in TESS observations and confirmed by ground-based data for \rx.
\cite{Sowicka2023} point out that \rx lies outside the theoretical blue edge of the GW Vir instability strip and raise the possibility of binarity or rotation.\\
A non-local thermodynamic equilibrium (non-LTE) spectral analysis of the optical spectrum of \rx obtained from \cite{Cowley1995} was carried out by \cite{inbook}. This analysis revealed an effective temperature of $\Teff = 180\,000  \pm 20\,000$\,K, a surface gravity of \(\log g\) = 7.5 ± 0.5, and He, C, O abundances of  0.68, 0.21, and 0.11 (mass fractions). It is currently speculated that primarily nitrogen-rich ($\approx$ 1\% N/He) PG\,1159 stars pulsate \citep{NitrogenPulsation}; however, since no nitrogen had been detected in the atmosphere, \cite{Werner+2004} pursued the search for neon because $^{14}$N is transformed into $^{22}$Ne through two $\alpha$ captures. 
Thus, the high-resolution optical spectra obtained with ESO's Very Large Telescope (VLT) were analysed using a similar non-LTE model setup: He, C, O, and Ne mass fraction abundances of 0.66, 0.21, 0.11, and 0.02, respectively. However, neither the \Ion{Ne}{7} 3644 \r{A} line nor the \Ion{Ne}{7} multiplet at 3850 - 3910 \r{A} could be found in the data.\\
Like for many hot white dwarfs, soft X-ray emission from the photosphere was detected from \rx \citep{Cowley1995}.
The very faint hard X-ray emission detected in \rx \citep{X-ray1} remains unexplained \citep{Xray2}.
\\
In this work we present a non-LTE spectral analysis based on new far-UV spectra obtained with the Cosmic Origins Spectrograph (COS) of the Hubble Space Telescope (HST), as well as archival spectra. We also present the first kinematic analysis of this star and discuss the implications of our findings for its evolutionary status.
\section{Observations} 
\label{sec:selction}
\rx was observed with various telescopes and instruments. A summary of all observations used in this work is provided in Table~\ref{tab:obs}, and properties from \textit{Gaia} are listed in Table~\ref{tab:properties}.
\begin{table}[t]
\centering
\caption{Properties of \rx from \textit{Gaia}, spectral analysis, and evolutionary track estimates. Abundances are given in mass fractions.}
\label{tab:properties}
\begin{tabular}{ll}
\hline
\hline
\noalign{\smallskip}
RA \,(J2000) [deg]& $\phantom{0}20.72372$ \\
\noalign{\smallskip}
Dec \,(J2000) [deg] & $-75.35420$  \\
\noalign{\smallskip}
$pmra $ \,[mas/yr]   & $\phantom{0}66.22${$\pm 0.04$}    \\
\noalign{\smallskip}
$pmdec$ \,[mas/yr]   & $-25.47${$\pm 0.04$}  \\
\noalign{\smallskip}
$ \varpi$  \,[mas] & 1.2{$\pm 0.04$} \\
\noalign{\smallskip}
$d$\,[pc]  & 834{$^{+27}_{-25}$}  \\
\noalign{\smallskip}
\hline
\noalign{\smallskip}
$E_{44-55}$ \,[mag]  & {$0.0345 \pm 0.0016$}  \\
\noalign{\smallskip}
$v \sin(i)$ \,[km/s]  & $50^{+20}_{-15}$  \\
\noalign{\smallskip}
\vrad$^{\mathrm{a}}$\,[km/s] & 110 $\pm$ 7\\
\noalign{\smallskip}
$v_{\mathrm{g}}$  \,[km/s] & $23 \pm 5$ \\
\noalign{\smallskip}
$T_{\mathrm{eff}}$ [K] & $175\,000^{+20000}_{-10000}$ \\
\noalign{\smallskip}
$\log g $& 7.7 $\pm$ 0.2 \\
\noalign{\smallskip}
$H$ & < 0.18 \\
\noalign{\smallskip}
$\mathit{He}$& 0.626 $\pm$ 0.003 \\
\noalign{\smallskip}
$C $& 0.372 $\pm$ 0.005 \\
\noalign{\smallskip}
$N$& < 0.00015 \\
\noalign{\smallskip}
$O$& 0.0027 $\pm$ 0.0004 \\
\noalign{\smallskip}
$\mathit{Si}$ & < 0.003 \\
\noalign{\smallskip}
$M_\mathrm{spec}$ \,[\Msol]
&$1.8^{+1.1}_{-0.7}$  \\
\noalign{\smallskip}
$R_\mathrm{spec}$ \,[\Rsol]
&$0.0312^{+0.0016}_{-0.0017}$ \\
\noalign{\smallskip}
$L_\mathrm{spec}$ \,[\Lsol]
&$840^{+440}_{-210}$  \\
\noalign{\smallskip}
\hline
\noalign{\smallskip}
$M_{\rm merger}$ \,[\Msol] & $0.78^{+0.04}_{-0.08}$  \\
\noalign{\smallskip}
$R_{\rm merger}$ \,[\Rsol] & $0.021^{+0.008}_{-0.004}$ \\
\noalign{\smallskip}
$L_{\rm merger}$ \,[\Lsol] & $373^{+332}_{-166}$ \\
\noalign{\smallskip}
$M_{\rm VLTP}$ \,[\Msol]&  $0.70^{+0.08}_{-0.06}$ \\
\noalign{\smallskip}
$R_{\rm VLTP}$ \,[\Rsol] & $0.020^{+0.009}_{-0.004}$ \\
\noalign{\smallskip}
$L_{\rm VLTP}$ \,[\Lsol]&$305^{+321}_{-146}$  \\
\noalign{\smallskip}
\hline
\end{tabular}
\tablefoot{~\\
\tablefoottext{a}{Corrected for gravitational redshift.}
}
\end{table}
\subsection{Archival spectra} 
Two high-resolution optical spectra were recorded at the VLT using the high-resolution UV-Visual Echelle Spectrograph (UVES) as part of the ESO Supernovae Ia Progenitor Survey (SPY; PI: Napiwotzki, ProgID: 167.D-0407). UVES operated in dichroic mode (Dichroic 1) with central wavelengths of 3900 \r{A} and 5640 \r{A}, and the resolving power is \(\lambda / \Delta \lambda = 19\,540\) for CD2 and \(\lambda / \Delta \lambda = 18\,770\) for CD3. The first UVES spectrum was recorded on February 9, 2001, with a S/N $=13$, and the second on September 14, 2001, with a S/N $=28$. The UVES spectra were downloaded from the ESO science archive. To determine the atmospheric parameters, we employed the co-added UVES spectrum. The spectra show weak lines of \Ion{He}{2}, \Ion{C}{4}, and \Ion{O}{6}.\\
An additional spectrum was obtained with the Space Telescope Imaging Spectrograph (STIS) on the Hubble Space Telescope (HST), covering the UV to near-infrared range (2000 \r{A} - 10\,000 \r{A}). It combines three spectra from low-resolution gratings G230LB, G430L, and G750L, which were averaged in the overlap regions (2990 – 3060 \r{A} and 5500 – 5650 \r{A}). The final spectrum has a resolving power of $\lambda / \Delta \lambda = 1000$ and includes interstellar dust correction\footnote{For more details on the reduction process, refer to: \url{http://archive.stsci.edu/pub/hlsp/stisngsl/aaareadme.pdf}}.
\subsection{New HST/COS UV spectrum} 
\begin{figure*}[ht]
  \centering
  \includegraphics[angle=0, width=\textwidth]{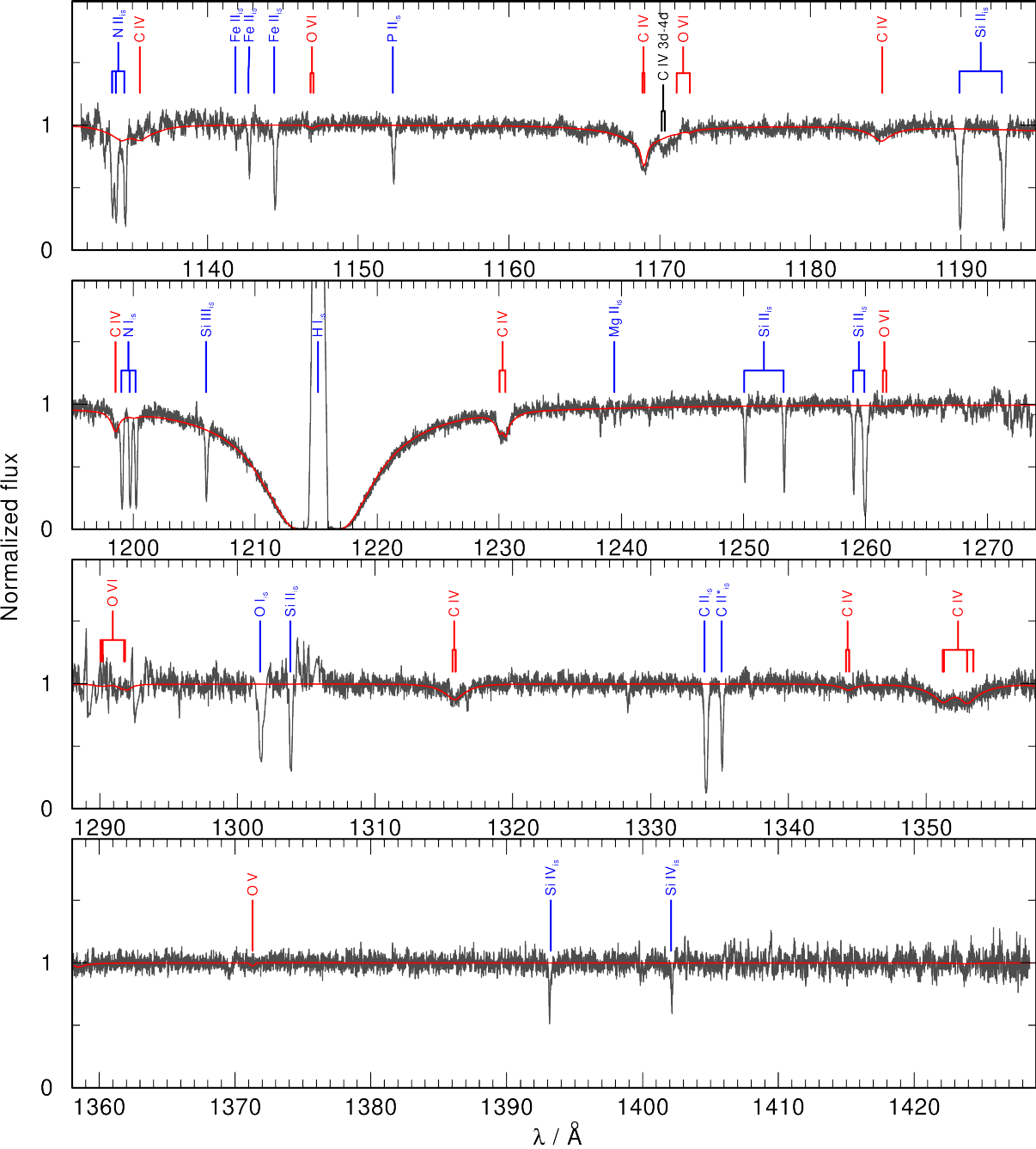}
  \caption{COS spectrum of \rx (in grey) with our best-fit photospheric model overlaid in red (175\,000\,K and \ensuremath{\log g = 7.7}; other parameters are listed in Table~\ref{tab:properties}). The most prominent photospheric lines are highlighted in red, and interstellar lines are marked in blue. In black, the forbidden \IonT{C}{4} 3d–4d transitions at 1170.13\,\AA\ and 1170.33\,\AA\ are highlighted. To match the observed spectra, the absorption feature of \IonT{H}{1} Lyman-alpha was additionally modelled with a column density of \ensuremath{n_{\text{HI}} = 4.5 \times 10^{20} \, \text{cm}^{-2}}.  }
  \label{fig:COS}
\end{figure*}

The far-UV spectrum was obtained with HST/COS
on March 18, 2023, during Cycle 30, as part of the ‘A Treasury FUV Survey of the Hottest White Dwarfs’ observing program (Proposal ID 17112, PI: Reindl). The observation employed the G130M grating with a central wavelength of 1291 \r{A} and a resolving power of $\lambda$ / $\Delta \lambda$ = 15\,000 - 19\,000.\\
The HST/COS spectrum is shown in Fig.~\ref{fig:COS}. Given the high effective temperature of the star, only a few weak lines of carbon and oxygen are visible. Yet, a remarkable feature in this spectrum is the clearly rotationally broadened \Ionww{C}{4}{1230.0, 1230.5} lines. Such significant rotational broadening has never been observed unambiguously in any other PG\,1159 star, nor in any of the more than 100 hot white dwarfs in the programme. We also highlight the presence of the forbidden \Ion{C}{4} 3d–4d transition at 1170.13\,\AA\ and 1170.33\,\AA.
\cite{forbidden} reported the discovery of forbidden \Ion{C}{4} transitions in several PG\,1159 stars and DO white dwarfs.
\section{Spectral analysis}
\label{sec:spec}
For the model calculations, we employed the T{\"u}bingen non-LTE model-atmosphere package
(\emph{TMAP}\footnote{http://astro.uni-tuebingen.de/\textasciitilde TMAP}; \citealt{werner+2003, rauchdeetjen2003, Tmap2012}), which allows the calculation of plane-parallel, non-LTE, fully metal-line blanketed model atmospheres in radiative and hydrostatic equilibrium. Subsequently, while keeping the atmospheric structure fixed, a line formation iteration was also performed to account for more detailed model atoms. Model atoms were taken from the T{\"u}bingen Model Atom Database\footnote{http://astro.uni-tuebingen.de/\textasciitilde TMAD}(TMAD), and the statistics of the model atoms used in our model atmosphere calculations are listed in Table~\ref{tab:Atoms}.
Using the results from \cite{inbook} as a starting point, we computed a model grid that considers abundances from He, C, and O. We computed a refined grid covering \(\Teff\) = \{160, 180, 200, 220\} kK, \(\log g\) = \{7.5, 8.0, 8.5\}, \(C\) = \{0.342, 0.424, 0.495\}, and \(O\) = \{0.00213, 0.00853, 0.0335\} (mass fractions). The best fit was determined using a $\chi^2$
minimization\footnote{These fits were performed in the ISIS framework \citep{2000ASPC..216..591H}.} to the archival spectra and the new COS UV spectrum. Systematic errors were estimated by considering the sensitivity of the fit results to different spectral regions. The determined parameters are listed in Table~\ref{tab:properties}.
\begin{table}[t]
\caption{\label{tab:Atoms}Model atoms for atmospheric structure calculations.}
\centering
\begin{tabular}{llrrr}
    \hline\hline
    Atom & Ion & non-LTE & LTE & Lines \\
    \hline
    \noalign{\smallskip}
    H & I & 5 & 27 & 10 \\
      & II & 1 & 0 & 0 \\
    \noalign{\smallskip}
    He & I & 5 & 98 & 3 \\
       & II & 15 & 17 & 105 \\
       & III & 1 & 0 & 0 \\
    \noalign{\smallskip}
    C & III & 6 & 99 & 4 \\
      & IV & 54 & 4 & 295 \\
      & V & 1 & 0 & 0 \\
    \noalign{\smallskip}
    O & IV & 8 (83) & 89 (14) & 9 (637) \\
      & V & 6 (104) & 128 (30) & 6 (751) \\
      & VI & 27 (54) & 35 (8) & 95 (291) \\
      & VII & 1 & 0 & 0 \\
    \hline
\end{tabular}
\tablefoot{
The first column lists the added elements, the second their ionization levels. The third and fourth columns show the number of non-LTE and LTE transitions used in the atmospheric structure calculation, respectively, while the fifth column indicates the calculated lines. Values in brackets are given when a different number of levels and lines was used for the line formation calculation.
}
\end{table}
\subsection{Radial velocity}
\label{sec:rv}
From the COS spectrum, we determine a radial velocity of $133\pm 2 \,\text{km/s}$. This is consistent with the radial velocity derived from the co-added UVES spectrum $v_{\mathrm{rad}}=133\pm2$\,km/s. 
When considering the two UVES spectra separately, we found no evidence of radial velocity variability.
\subsection{Projected rotational velocity}
\label{sec:vsini}
The projected rotational velocity was difficult to determine due to the lack of narrow metal lines. A careful analysis revealed that the \Ion{C}{4} lines between 1350\,\AA\ and 1355\,\AA\ tend to indicate an unrealistically high value, inconsistent with the projected rotational velocity obtained from the other lines. Excluding these lines from the fitting process results in a lower projected rotational velocity of 50\,km/s without compromising the overall fit quality.
To estimate the uncertainties, we examined the impact of the projected rotational velocity on the \Ionww{C}{4}{1230.0, 1230.5} lines. 
We note that the depth of the lines is not perfectly matched, an issue that has already occurred in previous analyses of hot white dwarfs \citep{WernerRauch2015, Wassermann+2010, badewanneTiefe}. Nevertheless, it becomes clear that for projected rotational velocities lower than 35\,km/s, the lines should appear more distinctly separated, and beyond 70 km/s, the lines are expected to appear significantly smoother than they do in the spectrum.
\subsection{Magnetic field}
\label{sec:magnetic}
Since Zeeman splitting scales with $\lambda^2$, the high resolution optical spectrum allows the best constraints on the upper limit of the magnetic field strength. The emission line cores of \Ion{He}{2}, \Ion{C}{4}, and \Ion{O}{6} do not exhibit splitting due to the presence of a magnetic field. Since rotation acts as the main broadening mechanism, we defined the detection threshold for the splitting between an $\sigma$ component and the unshifted central $\pi$ component as $\delta \lambda = 1.708 \,v \, \sin(i) \, \lambda_0 / c$ \citep{2005oasp.book.....G}. The corresponding upper limit for the magnetic field strength, $B$, that would produce such a splitting in the linear regime can be estimated following Eq. (10) from \cite{Wickramasinghe+2000}:
$$B[\mathrm MG] = \frac{\delta \lambda}{7.9\,\left(\frac{\lambda_0}{4101\,\AA}\right)^2}\,.$$
Using \Ionw{O}{6}{5390} as a reference line and the projected rotational velocity from \se{sec:vsini} allows us to put an upper limit of $\approx 110$\,kG on the magnetic field strength of \rx.
\subsection{Atmospheric parameters}
\begin{figure*}[!htbp]
  \centering
  \includegraphics[width=\textwidth]{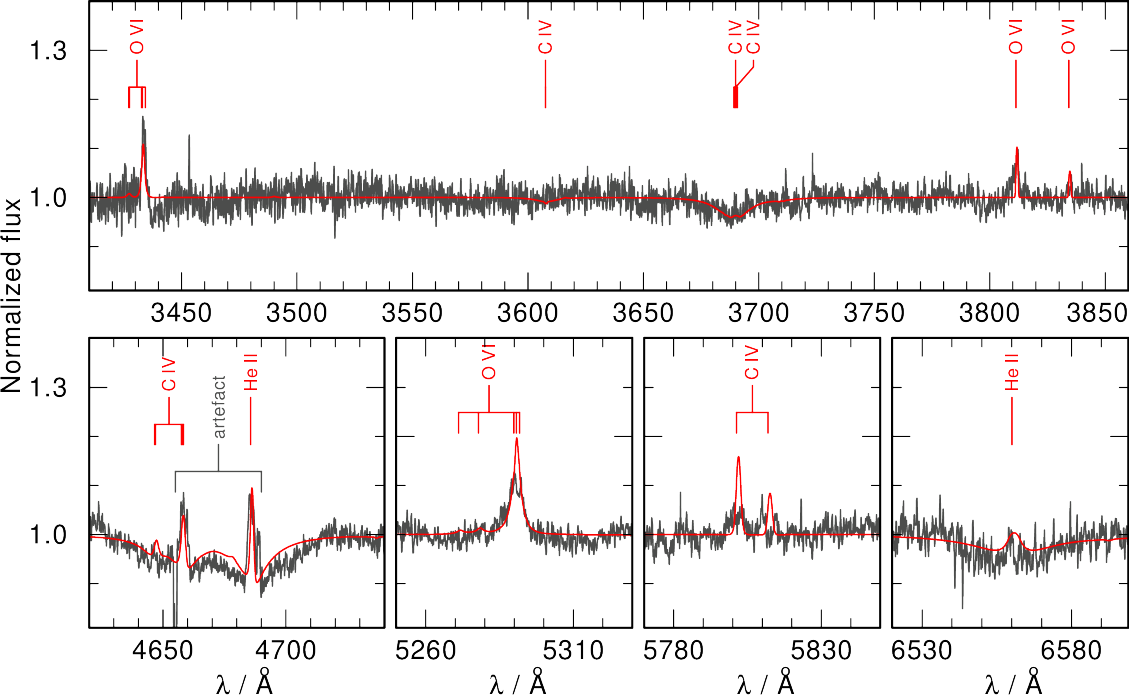}
  \caption{Relevant parts of the co-added UVES spectrum compared with our best-fit model (in red; 175\,000 K and \(\log g\) = 7.7; other parameters are listed in Table~\ref{tab:properties}). To reduce the noise the observation was convolved with a Gaussian (full width at half maximum $=0.2\AA$). Photospheric lines are marked in red, artefacts in grey.}
  \label{fig:UVES}
\end{figure*}

\begin{figure*}[htbp]
  \centering
  \includegraphics[angle=0, width=\textwidth]{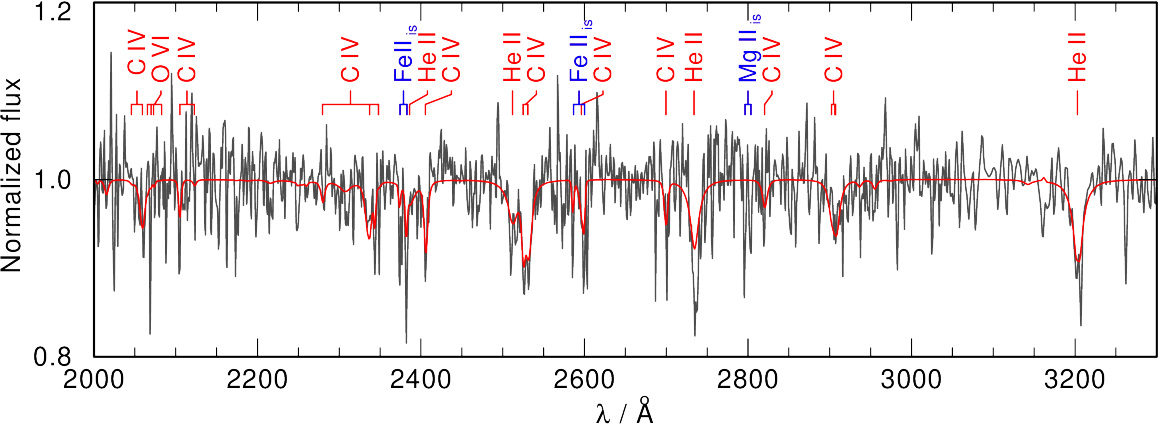}
  \caption{Important parts of the low-resolution STIS spectrum, with our best-fit model (in red; 175\,000 K and \(\log g\) = 7.7; other parameters are listed in Table \ref{tab:properties}). The interstellar lines are marked in blue.
}
  \label{fig:UVES}
\end{figure*}
A lower limit on \Teff of 165\,kK can already be concluded from the absence of \Ionw{O}{5}{1371.3} in the observed COS spectrum. The $\chi^2$ fit yielded \Teffw{175^{+20}_{-10}}, which is only slightly lower than the previously reported value of $180\pm20$\,kK by \cite{inbook}.
Additionally, we determine a surface gravity of \loggw{7.7\pm0.2}, which is somewhat higher but still in agreement with the value of \loggw{7.5\pm0.5} reported by \cite{inbook}. The upper limit of the temperature, as well as the uncertainty of $\pm0.2$ in \logg, are estimated systematic uncertainties.
For the He/C/O abundances, we determined values of 0.626/0.372/0.0027 (mass fractions), with the statistical uncertainties listed in Table~\ref{tab:properties}. Our best fits are shown in Figs.~\ref{fig:COS}-\ref{fig:UVES}.\\
The values for C and O differ from the values C/O $=0.21/0.11$ reported by \cite{inbook}.
This discrepancy arises from the uncertainty in determining the oxygen abundance from optical spectra.
In the case of \rx, the emission feature at 5293\,\AA\ was fitted in the previous work; however, its emission strength is also temperature- and gravity-dependent. At high oxygen abundances, this feature is expected to be accompanied by shallow absorption wings, which were not detectable in the relatively low-quality optical spectrum used in that study. Our much better spectrum reveals no such absorption wings, supporting our low value for the oxygen abundance.\\
We note that the \Ionww{C}{4}{5801, 5812} lines come out too strong in our model compared to the observation. This is a common issue observed among PG\,1159 stars, and DO-type white dwarfs. The C abundance derived from \Ionww{C}{4}{5801, 5012} appears to be systematically lower compared to what is derived from \Ionww{C}{4}{4646-4660}, and it was concluded that the likely origin of this issue could be an inaccuracy in the atomic data (see \citealt{Reindl+2018a} and references therein).\\
Potentially lines of \Ionw{Si}{5}{1251.4}, \Ionw{Si}{6}{1229.0}, and \Ionw{Si}{7}{1236.0} can be detected in spectra of the hottest white dwarfs \citep{WernerCoolPG1159,WernerRauch2015}. These lines are not visible in the spectrum of \rx, but in order to derive an upper limit for Si, we also included Si in our best-fitting model. We find that for $\text{Si} \geq0.003$ (by mass), the Si lines in our models become too strong.
In the same way, we were able to determine an upper limit for nitrogen (N < $0.00015$) based on the absence of the \Ionww{N}{5}{1238.8, 1242.8} resonance lines.
\\
Finally, we also investigated the possibility that \rx could be a hybrid PG\,1159 star. For this, we added H at the expense of He to our models. We find that for H > 0.18 (mass fraction), the \Ion{He}{2} Fowler lines that are visible in the near-UV part of the HST/STIS spectrum become too weak and therefore we determine $\text{H}=0.18$ (mass fraction) as an upper limit.
\section{Mass, luminosity, and radius}
\label{sec:mass}
\begin{figure*}[t]
    \centering
    \begin{minipage}{0.48\textwidth}
        \centering
        \vspace{0pt}  
        \includegraphics[width=\textwidth]{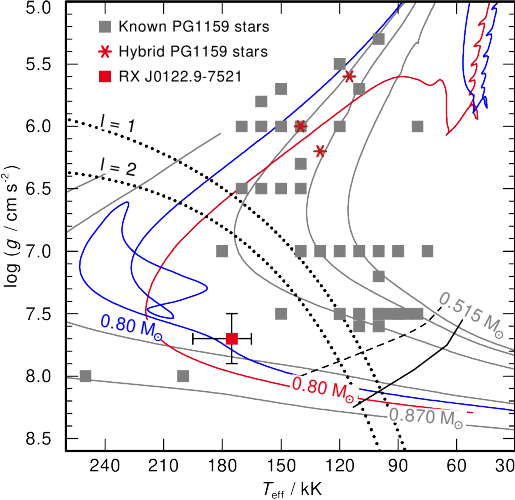}
        \caption{Position of \rx on the \(\log g\)-\(T_{\text{eff}}\) diagram, along with the other known PG\,1159 stars. VLTP evolutionary tracks from \cite{Althaus_2009} are depicted in grey and represent masses of 0.515, 0.530, 0.584, 0.741, and 0.870 $M_{\odot}$. The red track represents the potential evolutionary path of a merger product of a $0.45 M_{\odot}$ helium-core white dwarf and a $0.35 M_{\odot}$ carbon-oxygen-core white dwarf and was calculated by \cite{MillerB}. The blue line represents the merger of two He white dwarfs with a total mass of 0.8 $M_{\odot}$ \citep{Zhang+2012a}. The solid black line shows the wind limit, and the dashed black line shows the wind limit when the mass-loss rate is reduced by a factor of 10 \citep{UnglaubBues2000}. The blue edges of instability are indicated by the two dotted black lines, corresponding to the unstable g-modes l=1 and l=2, respectively \citep{2006A&A...458..259C}.}
        \label{fig:Kiel} 
    \end{minipage}
    \hfill
    \begin{minipage}{0.48\textwidth}
        \centering
        \vspace{-30pt} 
        \includegraphics[width=\textwidth]{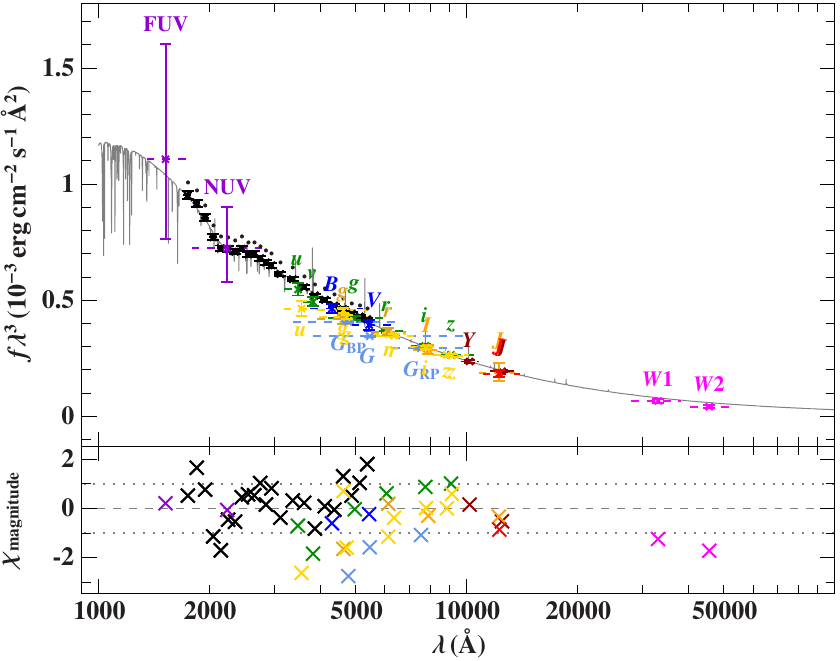}
        \vfill
        \caption{Fluxes averaged over filters, based on observed magnitudes, shown in different colours. Dashed horizontal lines indicate the full width at the tenth maximum for each filter passband, with the grey line representing the best-fitting model. Lower part: Uncertainty-weighted residuals. The colour coding for the various photometric systems is as follows: GALEX in violet \citep{Bianchi_2017}, 
    2MASS in red \citep{2MASS}, 
    DELVE DR2 in yellow \citep{DELVEDR2},
    DENIS in orange \citep{DENIS},
    \textit{Gaia} in light blue \citep{brown2021gaia},
    Johnson in blue \citep{Henden2015},
    SDSS in gold \citep{Alam_2015},
    SMASH in yellow \citep{SMASH},
    Skymapper in green \citep{Wolf2018},
    VISTA in maroon \citep{VISTA},
    WISE in magenta \citep{Schlafly2019}, and in black the STIS spectra G230LB | G430L | G750L with the box filters: [1000-3000\,\AA, step 100\,\AA], [3000-4500\,\AA, step 250\,\AA], and [4500-5500\,\AA, step 250\,\AA].}
        \label{fig:SED} 
    \end{minipage}
\end{figure*}

To estimate the mass, luminosity, and radius of \rx, we employed two methods. The first method relies on the position of \rx in the Kiel (\Teff$-$\logg) diagram, 
which is illustrated in Fig.~\ref{fig:Kiel}. We interpolated between evolutionary tracks, using the (V)LTP evolutionary tracks from \cite{Althaus_2009} and the double He-core white dwarf merger tracks from \cite{Zhang+2012a}. For the (V)LTP tracks, we derive a mass of $M_{\rm VLTP} = 0.70^{+0.08}_{-0.06}$\,\Msol. Via $R = \sqrt{G \cdot M/g}$ (where G is the gravitational constant) and calculate a radius of $R_{\rm VLTP} = 0.020^{+0.009}_{-0.004}$\,\Rsol. Using $L_{\rm VLTP}/L_\odot = (R_{\rm VLTP}/R_\odot)^2(T_\mathrm{eff}/T_{\mathrm{eff},\odot})^4$ we find a luminosity of $L_{\rm VLTP} = 305^{+321}_{-146}$\,\Lsol.
In the same way, from the merger tracks we derive the following values: $M_{\rm merger} = 0.78^{+0.04}_{-0.07}$\,\Msol, $R_{\rm merger} = 0.021^{+0.008}_{-0.004}$\,\Rsol\ and $L_{\rm merger} = 373^{+332}_{-166}$\,\Lsol.
The second method relies on the zero-point corrected \textit{Gaia} parallax distance, which accounts for its zero-point offset following \cite{Lindegren_SED}, and a fit of our best fitting model to the observed spectral energy distribution (SED), that takes into account the effect of interstellar reddening. For the SED fit, which is shown in Fig.~\ref{fig:SED}, we used the $\chi ^2$ fitting routine described in \cite{heber2018spectral} and \cite{Irrgang+2021}. The spectroscopic radius $R_{\rm spec}$ is given by $R_{\rm spec}=\theta /(2 \varpi)$, where $\varpi$ is the parallax and $\theta$ is the angular diameter, determined by photometric flux measurements. Interstellar extinction was modelled using the function given in \cite{2019ApJ...886..108F}, and we derive an extinction coefficient of R(55) = $2.8^{+0.4}_{-0.3}$.
From this, we determine a reddening of $E_\mathrm{44-55}=0.0345\pm0.0016$\,mag and a radius of $R_{\rm spec} = 0.0312^{+0.0016}_{-0.0017}$\,\Rsol. From this, in turn, we could calculate the spectroscopic mass, $M_{\rm spec}$, via $M_{\rm spec}/M_\odot=g/g_\odot (R_{\rm spec}/R_\odot)^2$, and we find a fairly high mass of $M_{\rm spec}=1.8^{+1.1}_{-0.7}$\,\Msol. 
The spectroscopic mass is inconsistent with the mass predicted by evolutionary models for merger scenarios, and also deviates from the mass estimate based on VLTP single-star evolution. However, it should be noted that for the hottest (pre-)white dwarfs, there is a recurring discrepancy between the spectroscopic and evolutionary masses, which is most likely due to unresolved systematic errors \citep{Ziegler+2012, Loebling+2020, Nicole_Blue, Reindl+2024}.
\\ 
We also investigated the impact of a hypothetical companion on the SED fit. For this, we chose to model an M dwarf, since we can directly rule out a brown dwarf, given that we would expect a significant reflection effect, which is not observed in any of our spectra.
To model the M dwarf, we used the empirical relationship from \cite{2015ApJ...804...64M} and applied their lower limit of Fe/He = -0.6. Our results indicate that we would be able to detect a companion until its temperature drops to \Teffwo{2900}, at which point the M dwarf would have a size of 0.1 $R_{\odot}$.
\section{Kinematic analysis}
\label{sect:kinematic}
\begin{figure}[H]
 \centering  
\includegraphics[width=0.9\linewidth]{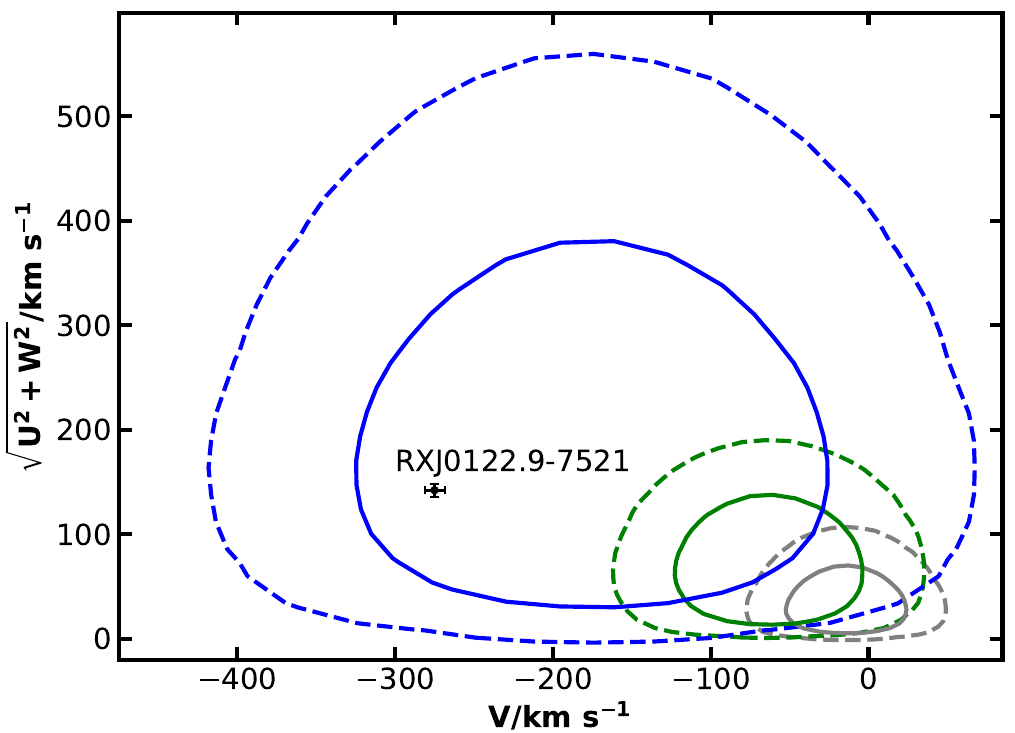}
 \caption{Toomre diagram for \rx. The lines represent the one- and two-sigma contours (thick and dashed line) -- blue for the Galactic halo, green for the thick disk, and grey for the thin disk -- derived from the U, V, and W velocity distributions provided by \cite{Kordopatis2011}.}
 \label{fig:Toomre}
\end{figure}

Given the high radial velocity, we further determined the space velocities using the \textit{Gaia} data, which are listed in Table~\ref{tab:obs}. 
The gravitational redshift correction was calculated using a mass of $M_{\text{VLTP}} = 0.70^{+0.08}_{-0.06} \, M_{\odot}$ and \(\log g\) = 7.7 $\pm$ 0.2, using the TGRED (Tübingen Gravitational REDshift) calculator\footnote{ 
\url{http://astro.uni-tuebingen.de/~TGRED/}}.
This yields a correction of $v_{\mathrm{g}} = 23 \pm 5 \,\text{km/s}$. Consequently, the velocity corrected for the gravitational redshift is $110 \pm 7 \,\text{km/s}$.
The Toomre plot clearly indicates that the calculated space velocities of \rx fall into a range distinctly associated with the galactic halo.
In addition, \rx shows a high eccentricity $e = 0.95 \pm 0.01$ and a low azimuthal action $L_z = -110 \pm 47$, which is consistent with halo kinematics \citep{Kordopatis2011}. Thus, we calculated the Galactic velocity components: \( U \), directed away from the Galactic centre; \( V \), aligned with the direction of Galactic rotation; and \( W \), pointing towards Galactic north. With these values, a Toomre plot was created as shown in Fig.~\ref{fig:Toomre}, which allows us to assign \rx to stars with similar dynamic properties. For this, the $1\sigma$ and $2\sigma$ contours for halo stars, thick disk stars, and thin disk stars from \cite{Kordopatis2011} were used.
The calculations were carried out using the Python package galpy (\citeyeartext{galpy}). Monte Carlo simulations were performed to account for uncertainties in the input parameters.

\section{Discussion}
\begin{figure*}[b!]
  \vspace*{\fill}
  \centering
  \includegraphics[width=0.80\textwidth]{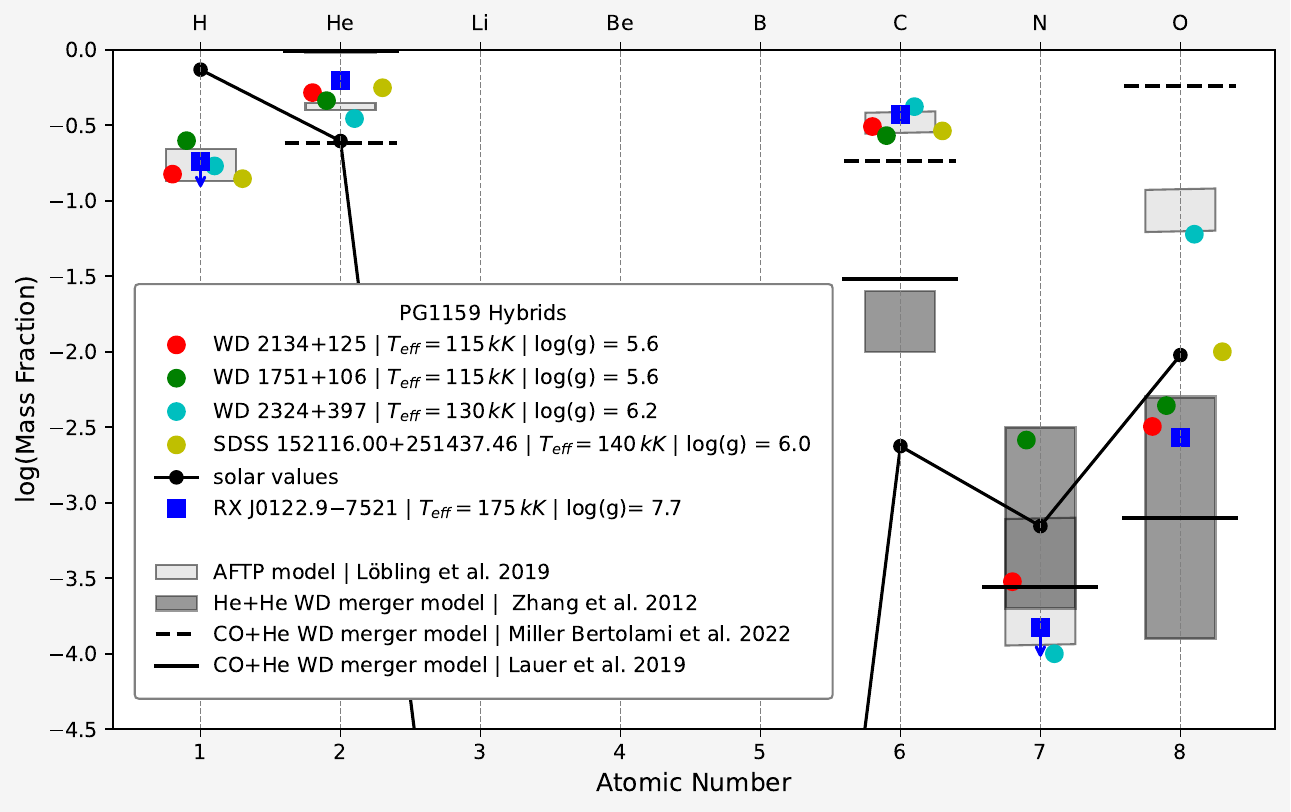}
  \caption{Element abundances found in the four hybrid PG\,1159 stars, along with the abundance of \rx determined in this work. For comparison with the theory, the AFTP model of \cite{PG1159LowO} and the merger models of \cite{Zhang+2012a}, \cite{MillerB}, and \cite{Lauer+2019} are included.} 
  \label{fig:Abundance}
  \vspace*{-0.5cm}
\end{figure*}

Our non-LTE spectral analysis, based on new UV and archival spectra of \rx,
revealed \Teffw{175^{+20}_{-10}}, \logg = 7.7 $\pm$ 0.2, and an astonishingly low O/C ratio of $7.3\times10^{-3}$ (by mass).
It is interesting to note that similarly low oxygen abundances were found in three hybrid PG\,1159 stars, namely \object{WD\,2134+125} [NGC 7094], \object{WD\,1751+106} [Abell 43], and \object{SDSS\,152116.00+251437.46} (\citeyeartext{PG1159LowO}; \citeyeartext{Werner+2014}). We compare the abundances of these stars to those of \rx in Fig.~\ref{fig:Abundance}. 
\\
Our spectral analysis suggests that a significant amount of hydrogen could be hidden. 
It is currently assumed that hybrid PG\,1159 stars are produced through an AFTP event. Models that account for this event, as described by \cite{AFTModel1}, would indeed lead to a comparable oxygen abundance. However, this finding is debated, as newer models by \cite{PG1159LowO} additionally include convective boundary mixing at the base of the pulse-driven convection zone, successfully reproducing H, He, C, and N values but failing to explain the low oxygen abundance.
\\
At the evolutionary stage of \rx, the gravitational settling of elements should not yet be the dominant effect; this is expected to occur only once the wind limit (indicated by the solid black line in Fig.~\ref{fig:Kiel}) is reached (\citeyeartext{UnglaubBues2000}).
A scenario with a mass-loss rate reduced by a factor of 10 was considered
(indicated by the dashed black line in Fig.~\ref{fig:Kiel}), which would lead to an earlier settling of the elements, but not early enough to affect \rx. Even if gravitational settling were to play a role (contrary to the predictions of the models), it would remain unclear why this effect is not also observed for carbon. 
\\
Our study has shown that other characteristics of \rx, in addition to the low oxygen abundance, are hard to reconcile with a canonical LTP scenario.
For instance, a period of 41\,min was reported by \cite{Sowicka2023}. 
Our analysis indicates that \rx exhibits neither radial velocity variability (\se{sec:rv}) nor an infrared excess (\se{sec:mass}), making it unlikely that the variability is caused by a companion. 
Additionally, there is no indication of pulsation since our analysis confirms that \rx is not located in the instability strip, as shown in Fig.~\ref{fig:Kiel}.
\\
We conclude that the 41~min period is indeed the rotational period of the star, which is supported by the clearly rotationally broadened \Ionww{C}{4}{1230.0, 1230.5} lines. The photometric variability might thus be related to a spot on the surface of the white dwarf, which in turn could be caused by a weak magnetic field \citep{Hermes+2017b, Hermes+2017c, UHE, Nicole_Blue}.
We note that the measured projected rotational velocity, $v \sin(i) = 50 ^{+20}_{-15} \, \text{km/s}$, falls within the expected range of $v_\text{rot} = 2\pi R / P \cdot \sin(i) = [28, 53] \, \text{km/s}$ for a period of $P = 41$ minutes, assuming radii between 0.016 and 0.030\,\Rsol\ (see \se{sec:mass}).\\
Interestingly, the 41~min period is significantly shorter than the typical rotation periods of the other seven PG\,1159 stars for which this value is known: these periods range from 5 to 37 hours, with an average value of \( P_{\text{rot}} = 22.5 \) h and a median value of \( P_{\text{rot}} = 28.0 \) h (\citeyeartext{C_rsico_2019}). 
Even among the group of slightly more evolved, very hot white dwarfs that exhibit the \Ion{He}{2} line problem\footnote{The poorly understood \Ion{He}{2} line problem affects about 10\% of all DO-type white dwarfs and refers to the challenge that the observed \Ion{He}{2} lines appear unusually deep and broad, and cannot be fitted with any current model \citep{Bedard+2020, UHE}} and/or absorption lines of
ultra-highly excited metals, \rx would still be considered an outlier. These hot white dwarfs have been found to represent a new group of variable stars, with their periods interpreted as the rotational periods \citep{UHE}. Their periods range from 3.5\,h to 2.9\,d \citep{UHE, Nicole_Blue}, which means that they rotate $5-100$ times more slowly than \rx. Once \rx has cooled to about 20\,kK, its radius will decrease by a factor of 2, and assuming conservation of angular momentum, the rotational period should decrease to approximately 10~min. Such a fast rotation is predicted by theoretical models for single white dwarfs that formed from double white dwarf mergers \citep{Schwab+2021} and is observed among DAQ white dwarfs \citep{Kilic+2024}, possibly hot DQ white dwarfs \citep{Williams+2016}, other types of (ultra-) massive white dwarfs \citep{Pshirkov+2020, Kilic+2021} and magnetic white dwarfs \citep{Kawka2020,  Ferrario+2020}.
\\
Kinematic analysis has identified \rx as a Galactic halo star. However, assuming single-star evolution and a Kiel mass of \( M_{\mathrm{VLTP}} = 0.70^{+0.08}_{-0.06} \, M_{\odot} \),
it deviates from the typical halo star mass. We would generally expect stars in this region to have masses of around \( M = 0.551 \pm 0.005 \, M_{\odot} \) (\citeyeartext{Kalirai2012}).
Furthermore, using the initial and final mass data provided by \cite{Althaus_2009}, we determined an initial mass of \(3.6 \pm 0.5 \, M_{\odot}\) based on $ M_{\mathrm{VLTP}}$. This corresponds to a main-sequence lifetime of $0.2\pm0.1$\,Gyr \citep{Renedo+2010}; \rx is thus far too young to have been born in the halo, whose age is estimated to be around 10--12 Gyr (\citeyeartext{10.1093/mnras/sty2755}; \citeyeartext{Guo_2016}).
The rapid rotation, combined with the too high mass for a halo star, suggests that the evolutionary history of \rx differs from previous assumptions about single-star evolution through LTPs and, instead, involved a double white dwarf merger.
\\
\cite{Werner+2022b} suggest that a fraction of the PG\,1159 stars might be successors of the CO-sdOs and result from the merger of a low-mass CO-core white dwarf and a more massive He-core white dwarf. However, this merger scenario cannot account for the very low oxygen abundance that was found for \rx. The He+CO white dwarf merger model from \cite{MillerB} predicts He, C, and mass fractions of O$=0.237$, $0.180$, and $0.578$, respectively (see also Fig.~\ref{fig:Abundance}, where the theoretically predicted values are shown as dashed lines), i.e. abundances similar to those found for the CO-sdOs, namely He, C, and O abundances of $\approx0.6$, $0.2$, and $0.2$ \cite{Werner+2022b}. 
While the model predicts a mass $M < 0.8 M_\odot$, smaller than our spectroscopic mass of $M_\mathrm{spec} = 1.8^{+1.1}_{-0.7} M_\odot$, the large uncertainty in the latter prevents any definitive conclusion.
Merger models in which a He-core white dwarf is accreted onto a CO white dwarf cannot explain the observed properties of \rx either. 
These models predict a mass of $M \leq 0.72 M_\odot$ \citep{Wu+2022}, 
significantly lower than the spectroscopic mass.
Furthermore, the CO+He white dwarf merger models predict too low abundances of C and O, and too high abundances of He and N (\citealt{Lauer+2019}, see also our Fig.~\ref{fig:Abundance}).
The double He-core white dwarf merger models of \cite{Zhang+2012a} predict a mass ($M = 0.78^{+0.04}_{-0.08} M_\odot$), comparable to those predicted by other merger scenarios, and again marginally inconsistent with our spectroscopic mass estimate.
In Fig.~\ref{fig:Abundance}, we show the abundances for C, N, and O as predicted by the composite merger model from \cite{Zhang+2012a} for $M=0.8$\,\Msol\ and $Z=0.02, 0.001$ (dark grey boxes). It becomes clear that while this model can reproduce the low oxygen abundance, it fails to account for the high C abundance and the upper limit of N in \rx.
In general, the resulting surface abundances of a merger product depend not only on the initial metallicity and core compositions of the interacting double white dwarf binary, but also on the mass ratio and accretion rates (e.g. \citealt{2014MNRAS.438...14D, Yu+2021}). This can potentially lead to a broad range of possible surface abundances in a merger remnant.
\\
In summary, several current evolutionary models fail to reproduce the observed abundance patterns of \rx. A single-star evolutionary scenario is in contradiction with the high mass and Galactic halo membership of \rx. The lack of evidence of a close binary and the rapid rotation favour a merger origin. Depending on the amount of H remaining in \rx, it might evolve into a massive, rapidly rotating DA or DB white dwarf.\\
We strongly encourage future evolutionary modelling that can explain the unique properties of \rx. In addition, future spectropolarimetric observations could help determine whether the observed photometric variability in \rx is connected to the presence of a weak magnetic field. This, in turn, could provide valuable insights into the evolution of magnetic fields in post-merger objects.
\begin{acknowledgements}
We thank Philipp Podsiadlowski for the helpful discussion.
N.M. is supported by the Deutsches Zentrum für Luft- und Raumfahrt (DLR) through grant 50 OR 2315.
N.R. is supported by the Deutsche Forschungsgemeinschaft (DFG) through grant RE3915/2-1.
M.D. was supported by the Deutsches Zentrum für Luft- und Raumfahrt (DLR) through grant 50-OR-2304.
This research is based on observations made with the NASA/ESA \textit{Hubble} Space Telescope obtained from the Space Telescope Science Institute, which is operated by the Association of Universities for Research in Astronomy, Inc., under NASA contract NAS 5–26555. These observations are associated with program 17112.
The SIMBAD database, managed by CDS in Strasbourg, France, was utilized for this research.
Data from the \textit{Gaia} mission, managed by the European Space Agency (ESA) (\url{https://www.cosmos.esa.int/gaia}), and processed through the \textit{Gaia} Data Processing and Analysis Consortium (DPAC, \url{https://www.cosmos.esa.int/web/gaia/dpac/consortium}) were used in this work. The TMAD tool (available at \url{http://astro.uni-tuebingen.de/~TMAD}), employed in this paper, was developed through the efforts of the German Astrophysical Virtual Observatory. 
Some of the data presented in this paper were obtained from the Mikulski Archive for Space Telescopes (MAST).
STScI is operated by the Association of Universities for Research in Astronomy, Inc., under NASA contract NAS5-26555. 
This research made use of NumPy \citep{2020Natur.585..357H}, Astropy \citep{2022ApJ...935..167A}, Matplotlib \citep{2007CSE.....9...90H}, SciPy \citep{2020NatMe..17..261V}, and galpy \citep{galpy}. This work made use of WRplot, which is available at: \url{https://www.astro.physik.uni-potsdam.de/~htodt/wrplot/index.html}.
\end{acknowledgements}
\bibliographystyle{aa}
\bibliography{BB}
\end{document}